\documentclass[12pt]{article}
\usepackage{amsmath,epsf,amssymb,latexsym,cite,shadow,eucal}
\usepackage[ps,dvips,matrix,arrow,frame,import,curve,color]{xy}

\newtheorem{theorem}{Theorem}

\newif\iffigs\figstrue

\DeclareFontFamily{U}{rsf}{}
\DeclareFontShape{U}{rsf}{m}{n}{
  <5> <6> rsfs5 <7> <8> <9> rsfs7 <10-> rsfs10}{}
\DeclareMathAlphabet\Scr{U}{rsf}{m}{n}

\def\O{\Scr{O}}
\def\C{{\mathbb C}}
\def\P{{\mathbb P}}

\def\Z{{\mathbb Z}}

\def\Hom{\operatorname{Hom}}

\def\Ext{\operatorname{Ext}}

\def\Spec{\operatorname{Spec}}
\def\Proj{\operatorname{Proj}}

\def\Sl{\operatorname{SL}}

\def\Cone{\operatorname{Cone}}

\def\p{\partial}

\def\CY{Calabi--Yau}
\def\LG{Landau--Ginzburg}

\def\cT{{\Scr T}}

\def\cF{{\Scr F}}

\def\DC{\mathbf{D}}

\def\mf#1{\mathfrak{#1}}

\def\DSing{\operatorname{\mathbf{D}^{\mathrm{gr}}_{\mathrm{Sg}}}}

\begin{document}

\begin{titlepage}
\begin{flushright}
DUKE-CGTP-06-04\\
hep-th/0610209\\
October 2006
\end{flushright}
\vspace{.5cm}
\begin{center}
\baselineskip=16pt
{\fontfamily{ptm}\selectfont\bfseries\huge
The Landau--Ginzburg to\\[3mm] Calabi--Yau Dictionary for D-Branes}\\[20mm]
{\bf\large  Paul S.~Aspinwall
 } \\[7mm]

{\small

Center for Geometry and Theoretical Physics, 
  Box 90318 \\ Duke University, 
 Durham, NC 27708-0318 \\ \vspace{6pt}

 }

\end{center}

\begin{center}
{\bf Abstract}
\end{center}
Based on work by Orlov, we give a precise recipe for mapping between
B-type D-branes in a Landau--Ginzburg orbifold model (or Gepner model)
and the corresponding large-radius \CY\ manifold. The D-branes in
Landau--Ginzburg theories correspond to matrix factorizations and the
D-branes on the \CY\ manifolds are objects in the derived category.
We give several examples including branes on quotient singularities
associated to weighted projective spaces. We are able to confirm
several conjectures and statements in the literature.  \vspace{2mm}
\vfill \hrule width 3.cm \vspace{1mm} {\footnotesize \noindent email:
  psa@cgtp.duke.edu}
\end{titlepage}

\vfil\break


\section{Introduction}    \label{s:intro}

B-type D-branes allow remarkable insight into the phase picture
\cite{W:phase,AGM:II} of \CY\ manifolds. Given a smooth \CY\ manifold
$X$, it is known that B-type D-branes are described by the
derived\footnote{All derived categories in this paper will be
  {\em bounded}.} category of coherent sheaves
\cite{Doug:DC,Doug:DICM,AL:DC,AD:Dstab}.  However, as one varies the
K\"ahler form on $X$, one can move into other phases where the \CY\
and its D-branes have another interpretation.

The most obvious ``other phase'' is the one studied originally in
\cite{W:phase}, namely the Landau--Ginzburg phase. Note that this
phase need not exist, but we will restrict attention here to cases
where it does. This Landau--Ginzburg theory also may, or may not, have
a Gepner model description \cite{Gep:} corresponding to the
construction of \cite{VW:}.

Certain D-branes in Gepner models were studied in
\cite{RS:DGep,BDLR:Dq}. A more general picture of D-branes in terms of
\LG\ theories was then given in
\cite{KL:Mfac,Kapustin:2003rc,Brunner:2003dc,Lazaroiu:2003zi}. Suppose
$X$ is described by a hypersurface $W=0$ in a (weighted) projective
space. The corresponding \LG\ theory then has a superpotential $W$.
B-type D-branes in this theory correspond to matrix factorizations:
\begin{equation}
  A.B = W.\textrm{id},
\end{equation}
where $A$ and $B$ are matrices (of arbitrary dimension). The
correspondence between Gepner model D-branes and such matrix
factorizations was described in \cite{AADia:GepD}.

Since the B-model is invariant to deformations of the K\"ahler form,
the category of (topological) B-type D-branes in the \LG\ phase must
be equivalent to the category of such branes in the large radius \CY\
phase. It is thus natural to ask how this correspondence works
exactly. Namely, given a matrix factorization, how does one find the
corresponding geometric D-brane on $X$? Or, given a geometric brane,
what is the corresponding matrix factorization?

There has been some progress in answering this question in several
examples \cite{AADia:GepD,Brunner:2004mt,DGJT:D,Hori:2004ja,HW:mfac,
Brunner:2005fv,Enger:2005jk} but no
systematic machine to convert one picture into another has been
described.

Orlov \cite{Orlov:mfc} recently proved the equivalence of the category
of matrix factorizations and the derived category of coherent sheaves
on the corresponding hypersurface. Orlov's proof is actually
constructive and allows for an explicit mapping between these
categories. In this paper we will demonstrate how this works.

The mathematics involved is quite technical but it seems to be very
naturally tied to the phase structure of $N=(2,2)$ theories. The
principal concepts involved are quotient triangulated categories and
semiorthogonal decompositions. The categories of D-branes in the \LG\
phase and \CY\ phases are both quotients of the same initial
category. The semiorthogonal decompositions with respect to these
quotients allows for an explicit map between the two categories.

In section \ref{s:cons} we will review the necessary material we
require from \cite{Orlov:mfc} and give the recipe for turning matrix
factorizations into complexes of sheaves and {\em vice versa}. In some
cases, in particular for single points and rational curves, the process
is relatively easy and we discuss this in section \ref{s:easy}. The 
Recknagel--Schomerus branes (or RS-branes) and their bound states are
discussed in section \ref{s:RS}. The case of D-branes on quotient
singularities is discussed in section \ref{s:quo} and we give our
concluding remarks in section \ref{s:disc}.


\section{Orlov's construction}  \label{s:cons}

In this section we review Orlov's construction \cite{Orlov:mfc} and
show how it can be used to explicitly map between matrix factorizations
and coherent sheaves. We refer to Orlov's papers
\cite{Orlov:LG,Orlov:mfc} for all the proofs of the assertions below.

One proceeds through a sequence of equivalences between triangulated
categories which we describe in turn. Let us begin with a graded
polynomial ring
\begin{equation}
  B = k[x_0,\ldots,x_{n-1}],
\end{equation}
where $k$ is a field (i.e., $\C$ for our purposes). One may choose to
give all the variables grade one, but we can also consider weighted
projective spaces by allowing arbitrary degrees. One then defines the
superpotential of the \LG\ theory
\begin{equation}
  W = f(x_0,\ldots,x_{n-1}),
\end{equation}
as a homogeneous polynomial of total degree $d$. We then have a
quotient ring
\begin{equation}
  A = \frac{B}{(W)}.
\end{equation}
This defines a hypersurface $X$ given by $W=0$ in a (weighted)
projective space. In the language of algebraic geometry, $X=\Proj A$.

\subsection{$\textrm{DGrB}(W)$}

The category of B-type D-branes in a \LG\ theory was described in
\cite{KL:TLG1}. Objects $\bar P$ are ordered pairs of free $B$-modules
of arbitrary but equal rank with maps between them going in each
direction:
\begin{equation}
  \bar P = \Bigl(
\xymatrix@1{
  P_1 \ar@<0.6mm>[r]^{p_1}&P_0\ar@<0.6mm>[l]^{p_0}
}\Bigr). \label{eq:Pdef}
\end{equation}
The two maps satisfy the matrix factorization condition
\begin{equation}
  p_0p_1 = p_1p_0 = W.\textrm{id}.   \label{eq:mfact}
\end{equation}

A map $f:\bar P\to\bar Q$ is simply a pair of maps $f_0:P_0\to Q_0$
and $f_1:P_1\to Q_1$ such that all squares commute. Such a map is
said to be null-homotopic if there are maps $s_0:P_0\to Q_1$ and
$s_1:P_1\to Q_0$ such that
\begin{equation}
  f_0 = s_1p_0 + q_1s_0,\quad f_1 = q_0s_1 + s_0p_1.
\end{equation}
The category of D-branes is given by the homotopy category obtained by
identifying morphisms with maps modulo null-homotopies.  The Hilbert
space of open strings in the topological B-model between two branes is
given by the space of morphisms in this category.  Note that if either
$p_0$ or $p_1$ is the identity map, it follows from this construction
that $\bar P$ is equivalent to 0 in this category.

For the \LG-\CY\ correspondence, one must orbifold the \LG\ theory by
$\Z_d$. It was noted in \cite{Wal:LGstab} that the effect of this
orbifolding is to put a well-defined grading structure on the above
category. That is, $P_0$ and $P_1$ are graded $B$-modules and one defines
$p_0$ to have degree $d$ and $p_1$ to have degree 0. Open strings are
maps of degree zero.

We denote this category of D-branes on a \LG\ orbifold
$\textrm{DGrB}(W)$. One can show that this category is a triangulated
category. In particular the shift functor $[1]$ is defined by
\begin{equation}
  \bar P[1] = \Bigl(
\xymatrix@1{
  P_0 \ar@<0.6mm>[r]^{p_0}&P_1(d)\ar@<0.6mm>[l]^{p_1}
}\Bigr),
\end{equation}
where $(d)$ denotes a shift in grading. That is, if $M$ is a graded
module then $(M(d)_n)=M_{d+n}$. Note in particular that
\begin{equation}
  \bar P[2] = \bar P(d). \label{eq:2=d}
\end{equation}

\subsection{$\DSing(A)$}

\def\grA{\hbox{gr-$A$}}

Orlov then shows that $\textrm{DGrB}(W)$ is equivalent to $\DSing(A)$
introduced in \cite{Orlov:LG} which is defined as follows. Let $\grA$
be the category of graded $A$-modules.\footnote{As $A$ is commutative,
  we need not distinguish between left and right actions.} Note that
the morphisms in $\grA$ are module homomorphisms {\em of degree zero}.
$\mathbf{D}(\grA)$ is then the bounded derived category of graded
$A$-modules.

Now let $\mf{Perf}(A)$ be the full subcategory of $\mathbf{D}(\grA)$
of objects which may be represented by finite-length complexes of free 
$A$-modules of finite rank.

One of the key ideas we require in this paper is the notion of a
quotient triangulated category. Given a triangulated category
$\mathsf{D}$ and full triangulated subcategory $\mathsf{N}$, we
define the quotient $\mathsf{D}/\mathsf{N}$ as follows. The objects in 
$\mathsf{D}/\mathsf{N}$ are the same as the objects in $\mathsf{D}$.
Consider the set of morphisms $\Sigma$ in $\mathsf{D}$ whose mapping
cones lie in $\mathsf{N}$. In other words $f:\mathsf{a}\to\mathsf{b}$
lies in $\Sigma$ if and only if we have a distinguished triangle
\begin{equation}
\xymatrix{
\mathsf{a}\ar[rr]^f&&\mathsf{b}\ar[dl]\\
&\mathsf{n}\ar[ul]|{[1]}
}
\end{equation}
where $\mathsf{n}$ is an object in $\mathsf{N}$. The morphisms in
$\mathsf{D}/\mathsf{N}$ are then defined by ``localizing'' on the set
$\Sigma$. That is, we invert the elements of $\Sigma$ in the same way
that quasi-isomorphisms are inverted in defining the derived category.

Note, in particular, that the zero map $0\to\mathsf{n}$ is in $\Sigma$
so that any element of $\mathsf{N}$ is isomorphic to zero in
$\mathsf{D}/\mathsf{N}$. 

We then define
\begin{equation}
   \DSing(A) = \frac{\mathbf{D}(\grA)}{\mf{Perf}(A)}.
\end{equation}

The correspondence between $\textrm{DGrB}(W)$ and $\DSing(A)$ may be
seen explicitly following a result of Eisenbud \cite{Eis:mf}. Consider
any $A$-module $M$ and compute a minimal free resolution. Eisenbud
showed that, if such a resolution has infinite length, then ultimately
it is periodic with period two. That is, we have an exact sequence:
\begin{equation}
\xymatrix@R=1mm{
  \cdots\ar[r]^{p_1}&P_0\ar[r]^{p_0}&P_1\ar[r]^{p_1}&P_0\ar[r]^{p_0}&P_1\ar[r]^{p_1}&\\
  &&\cdots\ar[r]&F^{-1}\ar[r]&F^0\ar[r]&M\ar[r]&0,
} \label{eq:resl1}
\end{equation}
where $P_0$, $P_1$ and $F^k$ are free $A$-modules, and $p_0$ and $p_1$
satisfy the matrix factorization condition (\ref{eq:mfact}).

So, given any $A$-module, we may map it to a matrix factorization by
computing a minimal free resolution. Clearly if two $A$-modules ``differ''
by an $A$-module with finite free resolution, they will produce the
same matrix factorization. Extending this to complexes gives the map
from $\DSing(A)$ to $\textrm{DGrB}(W)$. We will see examples of this later.

\def\DqgrA{\mathbf{D}(\hbox{qgr-$A$})}

\subsection{$\DqgrA$}

The next category we need to define is $\DqgrA$. An $A$-module is said
to be {\em torsion\/} if it is finite-dimensional as a vector space over
$k$. That is, it is annihilated by $x_i^N$ for any $i$ and
sufficiently large $N$. Let $\DqgrA$ be the quotient of
$\mathbf{D}(\grA)$ by the full subcategory given by complexes of
torsion modules.

It is a standard result due to Serre \cite{Serre:mp} that $\DqgrA$ is
equivalent to the derived category of coherent sheaves
$\DC(X)$.\footnote{See also exercise 5.9 in section II of
  \cite{Hartshorne:}. Also one should more correctly use the language
  of stacks for the weighted projective space case. We discuss this
  more in section \ref{s:quo}.} It is fairly easy to see why this should be so.
Start with the fact that $\P^{n-1}$ with homogeneous coordinates
$[x_0,\ldots,x_{n-1}]$ is constructed by $(\C^n -
(0,0,\ldots,0))/\C^*$. A standard construction in algebraic geometry
can be used to turn a module into a sheaf. We refer to
\cite{Hartshorne:} for details. In order to define a module on a
projective variety we demand that the module has a graded structure in
order to be compatible with the division by $\C^*$. Then note that any
module producing a sheaf supported at the origin $(0,0,\ldots,0)$ will
yield a trivial sheaf in $\P^{n-1}$. Ignoring such sheaves amounts to
taking a quotient by torsion sheaves.

So we have arrived at the statement that the equivalence between
\LG\ D-branes and large radius \CY\ D-branes is an equivalence between
$\DSing(A)$ and $\DqgrA$. That is, {\em we need to consider an equivalence
between two different quotients of the derived category of
graded $A$-modules.} This step is the only really substantial one in
understanding how to map between the two different kinds of D-branes.

\subsection{Semiorthogonal Decompositions}  \label{ss:decom}

The key concept in understanding the map between $\DSing(A)$ and $\DqgrA$ is
that of a {\em semiorthogonal decomposition}. Let $\mathsf{C}$ be a
triangulated category. We say that $\mathsf{C}=\langle\mathsf{A},
\mathsf{B}\rangle$ is a semiorthogonal decomposition\footnote{The
  semiorthogonal decompositions in this paper will all be ``weak'' in the
  sense of \cite{Orlov:mfc}.} of $\mathsf{C}$ if
the following three conditions are met:
\begin{enumerate}
\item $\mathsf{A}$ and $\mathsf{B}$ are full triangulated subcategories of
  $\mathsf{C}$.
\item For any object $\mathsf{c}$ in $\mathsf{C}$, there is a
  distinguished triangle
\begin{equation}
\xymatrix{
\mathsf{a}\ar[rr]|{[1]}&&\mathsf{b}\ar[dl]\\
&\mathsf{c}\ar[ul]
}
\end{equation}
in $\mathsf{C}$ where $\mathsf{a}$ is an object in $\mathsf{A}$ and
$\mathsf{b}$ is an object in $\mathsf{B}$.
\item $\Hom_\mathsf{C}(\mathsf{b},\mathsf{a})=0$ for any $\mathsf{a}$ in
  $\mathsf{A}$ and any $\mathsf{b}$ in
  $\mathsf{B}$.
\end{enumerate}

Fix an integer $i$ and let $\hbox{gr-$A$}_{\geq i}$ be the category of
graded $A$-modules $M$ such that $M_j$ is only nonzero if $j\geq i$.
Then $\DC(\hbox{gr-$A$}_{\geq i})$ is a full subcategory of
$\DC(\hbox{gr-$A$})$. Orlov then proved that there are two interesting
semiorthogonal decompositions
\begin{equation}
\begin{split}
\DC(\hbox{gr-$A$}_{\geq i}) &= \langle\mathcal{D}_i,\mathcal{S}_{\geq i}\rangle\\
\DC(\hbox{gr-$A$}_{\geq i}) &= \langle\mathcal{P}_{\geq i},\mathcal{T}_i\rangle.
\end{split}  \label{eq:decomp}
\end{equation}
The subcategories in this decomposition are defined as follows. Let
$\mathbf{k}$ be the $A$-module defined as the one-dimensional vector
space $k$ in grade 0 which is annihilated by $x_j$ for any $j$. Then
$\mathbf{k}(-e)$ is the corresponding one-dimensional space with grade
$e$.  We define $\mathcal{S}_{\geq i}$ as the triangulated subcategory of
$\DC(\hbox{gr-$A$}_{\geq i})$ generated\footnote{We will always assume
  {\em finitely\/} generated.} by $\mathbf{k}(-e)$ for $e\geq
i$.  As we will discuss further in section \ref{ss:tors}, the objects in
$\mathcal{S}_{\geq i}$ correspond to bounded complexes of torsion
modules in $\DC(\hbox{gr-$A$}_{\geq i})$.

$\mathcal{P}_{\geq i}$ is defined as the triangulated subcategory of
$\DC(\hbox{gr-$A$}_{\geq i})$ generated by $A(-e)$ for $e\geq
i$. Clearly objects in $\mathcal{P}_{\geq i}$ are the objects in
$\DC(\hbox{gr-$A$}_{\geq i})$ which have a finite length free
resolution. The subcategories $\mathcal{D}_i$ and $\mathcal{T}_i$ are
then defined from the decompositions (\ref{eq:decomp}).

We now claim that $\mathcal{T}_i$ is equivalent to $\DSing(A)$. To see
this note that the category of $A$-modules with finite free resolution
has a semiorthogonal decomposition $\langle\mathcal{P}_{\geq i},
\mathcal{P}_{<i}\rangle$, where $\mathcal{P}_{<i}$ is generated by
$A(-e)$ for $e<i$. One can then demonstrate the following equivalences:
\begin{equation}
\begin{split}
  \DSing(A) &= \frac{\DC(\hbox{gr-$A$})}{\langle\mathcal{P}_{\geq i},
\mathcal{P}_{<i}\rangle} \\
   &= \frac{\langle\DC(\hbox{gr-$A$}_{\geq i}),\mathcal{P}_{<i}\rangle}
      {\langle\mathcal{P}_{\geq i},
\mathcal{P}_{<i}\rangle} \\
   &= \frac{\DC(\hbox{gr-$A$}_{\geq i})}{\mathcal{P}_{\geq i}}\\
   &= \mathcal{T}_i
\end{split}  \label{eq:eq1}
\end{equation}

Let $\pi:\DC(\hbox{gr-$A$})\to\DqgrA$ be the quotient map. We now have
an explicit map 
\begin{equation}
\xymatrix@1{
  \Phi_i:\DSing(A)\ar[r]^-{\sim}&\mathcal{T}_i\ar[r]^-\pi&\DqgrA.
} \label{eq:Phi}
\end{equation}

Orlov shows that if $X$ is a \CY\ manifold, then $\Phi_i$ is an
equivalence and thus he proves that topological D-branes in a \LG\ theory
are the same as those on \CY\ manifold.

Similarly one can show that $\mathcal{D}_i$ is isomorphic to $\DqgrA$
and we have a map
\begin{equation}
\xymatrix@1{
  \Psi_i:\DqgrA\ar[r]^-{\sim}&\mathcal{D}_i\ar[r]^-\pi&\DSing(A),
} \label{eq:Psi}
\end{equation}
which takes geometric D-branes on a \CY\ manifold to matrix
factorizations.

\subsection{The explicit map}  \label{ss:map}

We now have a systematic method for computing the representation of a
D-brane as a complex of sheaves given the D-brane as a matrix
factorization. This map is essentially given above by (\ref{eq:eq1})
and (\ref{eq:Phi}). The steps are as follows:
\begin{enumerate}
\item Given a matrix factorization of the form (\ref{eq:Pdef}) we need
  to find any $A$-module $M$ with a resolution of the form
  (\ref{eq:resl1}). This could be done by setting $M$ equal to
  the cokernel of the map $p_1$, although we will not use this method below.
\item Use the semiorthogonal decomposition $\DC(\hbox{gr-$A$})
=\langle\DC(\hbox{gr-$A$}_{\geq i}),\mathcal{P}_{<i}\rangle$ to
decompose $M$ into a part $M'$ that lives in $\DC(\hbox{gr-$A$}_{\geq
  i})$ and a part, which we discard, that lives in $\mathcal{P}_{<i}$.
\item Use the semiorthogonal decomposition $\DC(\hbox{gr-$A$}_{\geq
    i})=\langle\mathcal{P}_{\geq i},\mathcal{T}_i\rangle$ to decompose
  $M'$ into a part $M''$ which lies in $\mathcal{T}_i$ and a part,
  which we discard, that lies in $\mathcal{P}_{\geq i}$.
\item The equivalence class of $M''$ in $\DqgrA=\DC(X)$ yields the
  desired complex of sheaves.
\end{enumerate}

Similarly we may start with a sheaf (or a complex of sheaves) on $X$
and find the corresponding matrix factorization with the following
steps:
{\renewcommand{\labelenumi}{$\theenumi'.$}
\begin{enumerate}
\item Given a coherent sheaf on $X$, we construct the corresponding
  graded $A$-module $M$ in the standard way.
\item Use the semiorthogonal decomposition $\DC(\hbox{gr-$A$})
=\langle\mathcal{S}_{<i},\DC(\hbox{gr-$A$}_{\geq i})\rangle$ to
decompose $M$ into a part $M'$ that lives in $\DC(\hbox{gr-$A$}_{\geq
  i})$ and a part, which we discard, that lives in $\mathcal{S}_{<i}$.
\item Use the semiorthogonal decomposition $\DC(\hbox{gr-$A$}_{\geq
    i})=\langle\mathcal{D}_{i},\mathcal{S}_{\geq i}\rangle$ to decompose
  $M'$ into a part $M''$ which lies in $\mathcal{D}_i$ and a part,
  which we discard, that lies in $\mathcal{S}_{\geq i}$.
\item Compute a minimal free resolution of $M''$ and use its
  asymptotic form to yield the matrix factorization.
\end{enumerate}}

\subsection{Monodromy}  \label{ss:mon}

It is important to realize that the maps $\Phi_i$ and $\Psi_i$ depend
on the choice of $i\in\Z$. The effect of changing $i$ is closely
related to the concept of monodromy and automorphisms of the triangulated
categories involved.

In both $\DqgrA$ and $\DSing(A)$ there is an automorphism of the
category generated by
\begin{equation}
  M \mapsto M(1).
\end{equation}
Clearly in terms of coherent sheaves, such a map corresponds to 
$\cF\mapsto\cF\otimes\O_X(1)$. This is well-known to correspond to
monodromy ``around the large radius limit''point, i.e., $B\mapsto
B+1$. In terms of \LG\ D-branes, Walcher's construction shows clearly
that this map corresponds to monodromy around the Gepner point.

It is therefore amusing to note that two completely different monodromies
are both represented by the same shift in the grading of the $A$-modules in
$\DC(\hbox{gr-$A$})$. This is possible, of course, because $\DqgrA$ and
$\DSing(A)$ are quite different quotients of $\DC(\hbox{gr-$A$})$.

The difference between monodromy around the Gepner point and monodromy
around the large radius limit can be determined in terms of monodromy
around the ``conifold'' point. Thus, the above observations should be
useful in verifying certain conjectures about monodromy and conifold
points as given in \cite{AKH:m0}, for example. We will not pursue this
issue here.

Anyway, it is clear from the definitions of $\Phi_i$ and $\Psi_i$ that
a shift in $i$ simply gives an automorphism of the category induced by
monodromy around the large radius limit, or the Gepner point. Such an
ambiguity will always be present in maps between the categories of
D-branes involved and we are required to make a choice.  {\em From now
  on, we will choose $i=0$.} We will see that this choice is
consistent with previous statements in the literature.


\section{Easy Cases}  \label{s:easy}

For the easiest correspondences between matrix factorizations and
sheaves it would be nice if we could evade steps $2'$ and
$3'$ in section \ref{ss:map}. That is, we could consider a free
resolution of a module associate to a sheaf and obtain the matrix
factorization immediately from its asymptotic form.

To this end, we will prove the following theorem
\begin{theorem} \label{th:easy}
  Let $\cF$ be the structure sheaf of a projectively normal subvariety
  of $X$ such that the cohomology groups
  $H^m(X,\cF(r))$ all vanish for $m>0$ and $r\geq0$. Let $M$ be the $A$-module
  associated with this sheaf. Then if $M$ has no negatively graded part,
  it lies in $\mathcal{D}_0$. Thus the matrix factorization
  is obtained directly from the free resolution of $M$.
\end{theorem}

To prove this we follow Orlov \cite{Orlov:mfc} again. Let
$\pi_0:\hbox{gr-$A$}_{\geq0}\to\hbox{qgr-$A$}$ be the map given by
inclusion into $\hbox{gr-$A$}$ followed by the natural projection.
Following \cite{AM:tor} Orlov shows that this functor has a right
adjoint $\omega_0$ which extends to the derived category
\begin{equation}
\mathbf{R}\omega_0:\DC(\hbox{gr-$A$}_{\geq0})\to \DC(\hbox{qgr-$A$}).
\end{equation}
Furthermore, the image of $\mathbf{R}\omega_0$ is precisely
$\mathcal{D}_0$.

Now assume $M\in \hbox{gr-$A$}_{\geq0}$. Then
\begin{equation}
\begin{split}
  M &= \bigoplus_{r=0}^\infty M_r\\
    &= \bigoplus_{r=0}^\infty \Hom_{\mathrm{gr-}A}(A,M(r)).
\end{split}
\end{equation}
So
\begin{equation}
\begin{split}
  \omega_0\pi_0 M &= \bigoplus_{r=0}^\infty
  \Hom_{\mathrm{gr-}A}(A,\omega_0\pi_0 M(r))\\
   &= \bigoplus_{r=0}^\infty
  \Hom_{\mathrm{qgr-}A}(\pi_0A,\pi_0 M(r))
\end{split}
\end{equation}
Using derived functors and the definition of sheaf
cohomology 
\begin{equation}
\begin{split}
R^j\omega_0\pi_0 M &= \bigoplus_{r=0}^\infty 
  R^j\Hom_{\mathrm{qgr-}A}(\pi_0A,\pi_0 M(r))\\
  &= \bigoplus_{r=0}^\infty H^j(X,\cF(r)),
\end{split} \label{eq:Rop}
\end{equation}
where $\cF$ is the sheaf on $X$ associated to $M$. Assuming $\cF$ has
no higher cohomology, we see that the complex representing
$\mathbf{R}\omega_0\pi_0M$ has cohomology only in position zero.
From (\ref{eq:Rop}) we obtain a homomorphism
\begin{equation}
 M\to \bigoplus_{r=0}^\infty H^0(X,\cF(r)).  \label{eq:nhom}
\end{equation}
Suppose this map is an isomorphism.  Then, the map
$M\to \mathbf{R}\omega_0\pi_0M$ is a quasi-isomorphism. We would then
have proven that $M$ lies in $\mathcal{D}_0$.

So when is the map in (\ref{eq:nhom}) an isomorphism?
This not true in general but exercise 5.14 in section II of
\cite{Hartshorne:} shows that it is true for the structure sheaf of a
``projectively normal'' variety. This technical condition will be true
for most simple examples and can be verified for the examples below.
Note, in particular, that a projectively normal variety must be
connected.

We may now use theorem \ref{th:easy} to prove some easy equivalences.

\subsection{Points}  \label{ss:points}

The obvious application is where $\cF$ is the skyscraper sheaf $\O_x$
of a point $x\in X$. There is no higher cohomology of a point!  The
analysis in this case is quite close to section 6.2 of
\cite{AADia:GepD}. Let $Y=\Proj B$ be the ambient projective space in
which $X$ is embedded.  Let $I_X$ be the homogeneous ideal in $B$
generated by homogeneous functions vanishing on $X$. Similarly $I_x$
is the homogeneous ideal associated to $x$. The statement that $x\in
X$ corresponds to the inclusion $I_x\supset I_X$.

We may always present a point as a complete intersection in $Y$, i.e., let
$I_x=(f_1,\ldots,f_{n-1})$. Clearly $I_X=(W)$. So we have a relation
\begin{equation}
  W = \sum_{i=1}^{n-1} f_i g_i, \label{eq:Wpt}
\end{equation}
for some $g_i\in B$.

The skyscraper sheaf $\O_x$ corresponds to the $A$-module
$M_x=A/(f_1,\ldots,f_{n-1})$. Consider taking a free resolution of this
module. By theorem \ref{th:easy} $M_x$ lies in $\mathcal{D}_0$ and, since
$\O_x$ is not a trivial sheaf, $M_x$ is not a trivial object in
$\DSing(A)$.

From what we have just said, this free resolution must be infinite,
and the asymptotic form will yield our desired matrix factorization.
Following \cite{AADia:GepD} we may write a matrix
\begin{equation}
  p = \left(\begin{matrix}0&p_0\\p_1&0\end{matrix}\right),
\end{equation}
mapping $P_0\oplus P_1$ on to itself. The matrix factorization
condition now becomes $p^2=W$.

The process of constructing the resolution begins with the functions
$f_i$ and analyzes the possible relations between them. Because we
have presented the point as a complete intersection or, in other
words, because $f_1,\ldots,f_{n-1}$ form a ``regular sequence'', there
will be no ``unexpected'' relations. We refer to chapter 17 of
\cite{Eis:CA} for more details.  Aside from the obvious Koszul
relations, we have (\ref{eq:Wpt}).  Because of this, the matrices
associated to the maps in the resolution have entries proportional to
any of the $f_i$'s or $g_i$'s. That is
\begin{equation}
  p = \sum_{i=1}^{n-1} \pi_i f_i+\bar\pi_i g_i,
\end{equation}
where $\pi_i$ and $\bar\pi_i$ are purely matrices of numbers.  The
matrix factorization condition then becomes
\begin{equation}
  \left\{\pi_i,\pi_j\right\}=0,\quad \left\{\bar\pi_i,\bar\pi_j\right\}=0,
  \quad \left\{\pi_i,\bar\pi_j\right\}=\delta_{ij}. \label{eq:Clif}
\end{equation}
That is, we have the Clifford algebra associated to the Hermitian
inner product of $\C^{n-1}$.

For a minimal resolution the matrices $\pi_i$ and $\bar\pi_i$ will
have dimension $2^{n-1}$ and thus our matrices $p_0$ and $p_1$ have
dimension $2^{n-2}$. It is easy to compute these matrices in any
example and the $8\times 8$ matrices associated to the quintic
threefold were given in \cite{AADia:GepD}.

If we consider a hypersurface $X$ in a weighted projective space we
need to be careful about the orbifold singularities. If a point lies
on a singularity then the matrix factorization will still work just as
above but the resulting D-brane will not have the same D-brane charge as
a point at a generic smooth point. Instead we will obtain a fractional
brane. We will discuss this more in section \ref{s:quo}.

It is worth pointing out that another obvious method of producing
points is not suitable for the method of theorem \ref{th:easy}.
Suppose we intersect $n-2$ generic hyperplanes in $\P^{n-1}$. The
intersection of this with $W=0$ yields $n$ points. Even though these
$n$ points are a complete intersection and their higher cohomology is
zero, they are not connected and so do not form a projectively normal
variety. In this case, (\ref{eq:nhom}) is not an isomorphism. Indeed,
a free resolution of the associated module in this case is finite and
does not produce a matrix factorization.

\subsection{Rational Curves} \label{ss:curves}

The next easiest case of a sheaf with trivial cohomology is $\O_C$,
the structure sheaf of a rational curve $C\cong\P^1$ which lies in
$X$.

If $C$ is a complete intersection in $Y$, the associated ideal is
generated by $n-2$ equations $f_1,\ldots,f_{n-2}$. The fact that this
curve lies in $X$ then yields
\begin{equation}
  W = \sum_{i=1}^{n-2} f_i g_i,
\end{equation}
for some polynomials $g_i$. The construction proceeds in exactly the
same way as the previous section. Now the Clifford algebra is
associated with $\C^{n-2}$ and our matrix factorization is for matrices
of dimension $2^{n-3}$.

For example, if $X$ is a 3-fold, the resulting $4\times 4$ matrix
factorization can be written
\begin{equation}
  p_0 = \left(\begin{matrix}f_1&-g_2&g_3&0\\
                           -f_2&-g_1&0&g_3\\
                           f_3&0&-g_1&g_2\\
                           0&f_3&f_2&f_1\end{matrix}\right),\quad
  p_1 = \left(\begin{matrix}g_1&-g_2&g_3&0\\
                           -f_2&-f_1&0&g_3\\
                           f_3&0&-f_1&g_2\\
                           0&f_3&f_2&g_1\end{matrix}\right)
\end{equation}

The case of the quintic threefold is well-studied. Any line (i.e.,
curve of degree one) must be the intersection of 3 linear equations
and is thus a complete intersection. Thus, the 2875 lines on a generic
quintic have $4\times 4$ matrix factorizations. Similarly, any quadric
curve lies in a $\P^2\subset\P^4$ and is thus is a complete
intersection. So the 609250 quadrics also have this simple matrix
factorization. 

However, cubic rational curves {\em cannot\/} be complete
intersections, although they are still projectively normal. If a
cubic curve is a complete intersection it would be a plane cubic and
thus an elliptic curve. Instead one uses so-called ``twisted cubics''.
So the cubic rational curves have a slightly more complicated
description in terms of matrix factorizations. Having said that, given
the presentation of any cubic curve, it is quite easy using a computer
package such as Macaulay to compute the corresponding \LG\ D-brane.
For example, consider the cubic rational curve
\begin{equation}
  x_1^2-x_0x_2=x_2^2-x_1x_3=x_1x_2-x_0x_3=x_4=0,
\end{equation}
in the smooth quintic 3-fold with defining equation
\begin{equation}
{{x}}_{0}^{3} {{x}}_{1} {{x}}_{{2}}+{{x}}_{1}^{3} {{x}}_{{2}}^{2}+
{{x}}_{1}^{2} {{x}}_{{2}}^{3}-{{x}}_{0} {{x}}_{{2}}^{4}-{{x}}_{0}^{4} 
{{x}}_{{3}}-{{x}}_{1}^{4} {{x}}_{{3}}+{{x}}_{{3}}^{4} {{x}}_{{4}}+{{x}}_{{4}}^{5}.
\end{equation}
Using Macaulay to compute a resolution in this case gives a $6\times
6$ matrix factorization.


\section{Recknagel--Schomerus Branes}  \label{s:RS}

Suppose we have an A-series Gepner model. That is, we have a
polynomial ring $B=k[x_0,\ldots,x_{n-1}]$ where the variables have
degrees $d_0,\ldots,d_{n-1}$ and a superpotential
\begin{equation}
  W = x_0^{l_0}+x_1^{l_1}+\ldots+x_{n-1}^{l_{n-1}},
\end{equation}
where $d=\sum_i d_i$ and the $l_i$'s are integral and equal to
$d/d_i$. Define $A=B/(W)$ as above.

View $W=0$ as an affine variety in $k^n$. The reason why $\DSing(A)$
is not trivial is that this variety has a singularity at the origin. In affine
language, the skyscraper sheaf at the origin corresponds to the module
$\mathbf{k}$ defined in section \ref{ss:decom}. That is, we have an
exact sequence:
\begin{equation}
\xymatrix@1{
  {\displaystyle\bigoplus_{i=0}^{n-1} A(-d_i)}\ar[rrr]^-{\left(
   \begin{smallmatrix}x_0&x_1&\ldots&x_{n-1}\end{smallmatrix}\right)}&&&
A\ar[r]&\mathbf{k}\ar[r]&0.
}
\end{equation}
We obtain a free resolution by continuing this exact sequence to the
left with free modules. This resolution is not finite because of the
singularity at the origin.

If we did this resolution in $B=k[x_0,\ldots,x_{n-1}]$ we would obtain
a finite resolution in terms of the usual Koszul resolution. All the
entries in the matrices of this resolution are linear in the
$x_i$'s. When we do the resolution over $A=B/(W)$, we obtain new
relations because of the obvious relation
\begin{equation}
  \sum_i x_i g_i = W,
\end{equation}
where $g_i=x_i^{l_i-1}$. We are now in a similar situation to section
\ref{ss:points}. The matrices in the infinite resolution now have
entries proportional to $x_i$ or $g_i$. Then
\begin{equation}
  p = \sum_i(x_i\pi_i + g_i\bar\pi_i),
\end{equation}
where $\pi_i$ and $\bar \pi_i$ are matrices of numbers. The matrix
factorization condition then produces the Clifford algebra
(\ref{eq:Clif}) associated to $\C^n$. Thus $p_0$ and $p_1$ are
matrices of dimension $2^{n-1}$. This is exactly the construction
used in \cite{AADia:GepD} to produce the tensor product D-branes of
the Gepner model. That is, we claim that $\mathbf{k}$ represents the
tensor product D-brane
\begin{equation}
  M_1(x_0)\otimes M_1(x_1)\otimes\ldots\otimes M_1(x_{n-1}),
\end{equation}
where $M_a(x)$ represents the matrix factorization of the minimal model
$x^a.x^{m-a} = x^m$.

It was demonstrated in \cite{AADia:GepD} that these tensor product
D-branes correspond to the RS D-branes of
\cite{RS:DGep} constructed directly from the Gepner model. An
RS D-brane is denoted by\footnote{Our subscript will be $r$ rather
  than the more conventional $M$ to be consistent with the notation in
  the rest of this paper.} $(L_0, L_1, \ldots
L_{n-1})_r$. The result of \cite{AADia:GepD} is that the D-brane
$(L_0, L_1, \ldots, L_{n-1})_0$ is equivalent to the matrix factorization
\begin{equation}
  M_{L_0+1}(x_0)\otimes M_{L_1+1}(x_1)\otimes\ldots\otimes M_{L_{n-1}+1}(x_{n-1}).
\end{equation}
Thus we have shown the correspondence of D-branes
\begin{equation}
  (0,0,\ldots,0)_0 \cong \mathbf{k}.
\end{equation}
Replacing $x_j$ by $x_j^{L_j+1}$ in the free resolution above, it is
easy to see that we have a more general correspondence
\begin{equation}
  (L_0, L_1, \ldots, L_{n-1})_0 \cong
  \frac{A}{(x_0^{L_0+1},x_1^{L_1+1},\ldots,
        x_{n-1}^{L_{n-1}+1})}.
\end{equation}
From Walcher's construction \cite{Wal:LGstab}, it is easy to see that
shifting $r$ in the RS D-branes is equivalent to
shifting the grading of the module. Thus we have the following
correspondence for an arbitrary RS D-brane:
\begin{equation}
  (L_0, L_1, \ldots, L_{n-1})_r \cong
  \frac{A}{(x_0^{L_0+1},x_1^{L_1+1},\ldots,
        x_{n-1}^{L_{n-1}+1})}(r).   \label{eq:RS}
\end{equation}
This completes step 1 of section \ref{ss:map}. We have found an
$A$-module whose free resolution corresponds to the desired matrix
factorization associated to any Recknagel--Schomerus D-brane.

\subsection{Type $(0,0,\ldots,0)_r$} \label{ss:00r}

In order to proceed let us focus on some specific D-branes. First
consider $(0,0,\ldots,0)_0\cong\mathbf{k}$. Clearly $\mathbf{k}$ has
only a degree 0 part with respect to the grading and so step 2 of
section \ref{ss:map} is trivial --- $\mathbf{k}$ is already in
$\DC(\hbox{gr-$A$}_{\geq0})$.

Suppose, for a moment, that step 3 were trivial too, i.e., that
$\mathbf{k}$ lay in $\mathcal{T}_0$. This would mean that our
resulting sheaf in the large radius interpretation would be the sheaf
on $X$ corresponding to the $A$-module $\mathbf{k}$. But $\mathbf{k}$
is a torsion sheaf and so, by Serre's construction, corresponds to the
zero sheaf. We know that $(0,0,\ldots,0)_0$ is a nontrivial D-brane
will results in a contradiction.

To perform step 3 we need a distinguished triangle
\begin{equation}
\xymatrix{p\ar[rr]|{[1]}&&t\ar[dl]\\&\mathbf{k}\ar[ul]} \label{eq:dtk}
\end{equation}
where $p$ is in $\mathcal{P}_{\geq 0}$ and $t$ is in
$\mathcal{T}_{0}$. From the definition of the semiorthogonal
decomposition, and the definition of $\mathcal{P}_{\geq 0}$, $t$ must
satisfy\footnote{From now on, any Hom or Ext appearing without a
  subscript will refer to the (derived) category of graded $A$-modules.}
\begin{equation}
  \Ext^m(t,A(r))=0, \quad\hbox{for all $m$ and for all $r\leq0$.}
     \label{eq:EE0}
\end{equation}
From the long exact sequence of Ext's associated to (\ref{eq:dtk}) we
obtain
\begin{equation}
  \Ext^m(p,A(r)) = \Ext^m(\mathbf{k},A(r)), 
\quad\hbox{for all $m$ and for all $r\leq0$.} \label{eq:EEE}
\end{equation}
Now, $A$ is an ``AS-Gorenstein'' ring since
\begin{equation}
  \mathbf{R}\Hom_A(\mathbf{k},A) = \mathbf{k}[-n+1]. \label{eq:ASG}
\end{equation}
This follows from Serre duality and the \CY\ condition on $X$. We
refer to lemma 2.11 in \cite{Orlov:mfc} for a proof of this
statement. It follows that
\begin{equation}
  \Ext^{n-1}(\mathbf{k},A) = k,
\end{equation}
with all the other Ext groups vanishing in (\ref{eq:EEE}). Thus we may
satisfy (\ref{eq:EEE}), and therefore (\ref{eq:EE0}), by setting $p=A[n-1]$.

The distinguished triangle (\ref{eq:dtk}) can also be used to perform
step 4. We need to project the object $t$ into in $\DqgrA=\DC(X)$. But
$\DqgrA$ is obtained by quotienting by torsion modules and
$\mathbf{k}$ is itself a torsion module. Thus, using the definition of
a quotient triangulated category (and shuffling around the $[1]$ in
(\ref{eq:dtk})), we see that $t$ is equivalent to $p[-1]=A[n-2]$ in
$\DqgrA$. It is basic to the construction of sheaves that $A$ as an
$A$-module corresponds to the structure sheaf $\O_X$. We the obtain
the correspondence
\fboxsep=5mm
\begin{equation}
  \fbox{$(0,0,\ldots,0)_0 \cong \O_X[n-2]$.} \label{eq:000}
\end{equation}
In the case of the quintic threefold, this correspondence has been
known (up to the shift) for some time \cite{BDLR:Dq}. Now we see this
result is completely general.

Next let us consider $(0,0,\ldots,0)_{-1}\cong\mathbf{k}(-1)$. This is
very similar to the above case of $(0,0,\ldots,0)_0$. Again the
projection to $\DC(\hbox{gr-$A$}_{\geq0})$ is trivial. Now, we require
$p$ to satisfy $\Ext^{n-1}(p,A(-1))=k$. This could be satisfied by
setting $p=A(-1)[n-1]$. However, we also require $\Ext^m(p,A)=0$ for
all $m$ and this choice would break this latter condition.

Recall that $\Hom_A(A,A)=A$ as an $A$-module. Thus, in the category of
graded $A$-modules, $\Hom(A,A(r))=A_r$ (i.e., the vector space of
elements of degree $r$ in the algebra $A$) with all higher $\Ext$'s
vanishing. We claim one should use the following complex for $p$:
\begin{equation}
\fboxsep=1pt
\xymatrix@1{A(-1)\save[]+<0mm,-5mm>*{\fbox{$\scriptstyle 1-n$}}\restore
  \ar[rr]^{\left(\begin{smallmatrix}x_0\\x_1\\
        \vdots\\x_{\alpha_1-1}\end{smallmatrix}\right)}&&
  A^{\oplus\alpha_1},}
\end{equation}
where $x_0,x_1,\ldots,x_{\alpha_1-1}$ are the variables of degree one
in the ring $B$. The $1-n$ in a box denotes the position in the
complex.  Applying $\Hom(-,A(-1))$ we see that $\Ext^{n-1}(p,A(-1))=k$.
Applying $\Hom(-,A)$ to this complex, the two terms in the complex
cancel and we obtain $\Ext^{n-1}(p,A)=0$ as desired. It is easy to
show that all other $\Ext$'s vanish as necessary justifying our claim.

Thus, in general, the D-brane $(0,0,\ldots,0)_{-1}$ corresponds to the
following complex of sheaves
\begin{equation}
\fboxsep=1pt
\xymatrix@1{\O_X(-1)\save[]+<0mm,-5mm>*{\fbox{$\scriptstyle 2-n$}}\restore
  \ar[rr]^{\left(\begin{smallmatrix}x_0\\x_1\\
        \vdots\\x_{\alpha_1-1}\end{smallmatrix}\right)}&&
  \O_X^{\oplus\alpha_1}.}
\end{equation}

Suppose all the variables in $B$ are of degree 1, i.e., we have a
non-weighted projective space. Then $d$, the degree of the
superpotential, equals $n$. On the corresponding $\P^{n-1}$ we have
the Euler exact sequence
\begin{equation}
\xymatrix@1{
  0\ar[r]&\O\ar[rr]^{\left(\begin{smallmatrix}x_0\\x_1\\
        \vdots\\x_{n-1}\end{smallmatrix}\right)}&&\O(1)^n\ar[r]&\cT\ar[r]&0,
}\end{equation}
where $\cT$ is the tangent sheaf of $\P^{n-1}$. It follows that
$(0,0,\ldots,0)_{-1}$ corresponds to $\cT(-1)[n-3]$ restricted to $X$.

Let $\Omega^k$ denote the $k$th exterior power of the cotangent bundle
of $\P^{n-1}$ restricted to $X$. Since $\Omega^{n-1}\cong\O_X(-n)$, it
follows that $\cT(-1)[n-3]=\Omega^{n-2}(n-1)[n-3]$.

In order to see the general picture (for arbitrary weights) for
$(0,\ldots,0)_r = \mathbf{k}(r)$ for any $r$, it is instructive to
consider the cases where $r>0$. Consider constructing a free resolution
$C\to\mathbf{k}$, where $C$ is an infinite complex of free
$A$-modules. This leads to a distinguished triangle in
$\DC(\hbox{gr-$A$})$:
\begin{equation}
\xymatrix{
  C(r)_{\geq0}\ar[rr]|{[1]}&&C(r)_{<0}\ar[dl]\\
  &\mathbf{k}(r)\ar[ul]} \label{eq:Cr}
\end{equation}
where $C(r)_{\geq0}$ is obtained from $C(r)$ by deleting from the
complex any $A(a)$ for $a>0$. Similarly $C(r)_{<0}$ is obtained from
$C(r)$ by deleting all the $A(a)$'s for $a\leq0$. Clearly $C(r)_{<0}$
is a finite length complex.

$C(r)_{\geq0}$ is trivially in $\DC(\hbox{gr-$A$}_{\geq0})$. Less
obviously, it is also in
$\mathcal{T}_0$ as can be seen as follows. Since
$\Ext^m(C(r)_{<0},A(s))=0$ for any $m$ and any $s\leq0$, it follows
from (\ref{eq:Cr}) that
$\Ext^m(C(r)_{\geq0},A(s))=\Ext^m(\mathbf{k}(r),A(s))$ for any $m$ and
any $s\leq0$. But, given that $r>0$, we see from (\ref{eq:ASG}) that
the latter is zero. It follows that $C(r)_{\geq0}$ is in
$\DC(\hbox{gr-$A$}_{\geq0})$. Finally we may use (\ref{eq:Cr}) again
to see that $C(r)_{\geq0}$ is equivalent to $C(r)_{<0}[1]$ in
$\DqgrA$.

We have therefore shown that the matrix factorization corresponding to
$\mathbf{k}(r)$ is equivalent to the finite complex $C(r)_{<0}[1]$
in $\DqgrA$.

To simply matters further, suppose $0<r<d$. In this case, the
complex $C(r)_{<0}$ obviously never contains any part of the free
resolution that gets to ``see'' the effects of the condition $W=0$.
This means that we may as well resolve $\mathbf{k}(r)$ as a
$B$-module, rather than an $A$-module. This is a straight-forward
finite Koszul resolution.

This makes computing the geometrical interpretation of the D-brane
$\mathbf{k}(r)$ very easy. In the case of a non-weighted projective
space (so $d=n$) we obtain
\begin{equation}
  \fbox{$(0,0,\ldots,0)_{r} \cong \Omega^{r-1}(r)[r]$,}
  \label{eq:RS00}
\end{equation}
for $0< r < n$. For the quintic threefold, putting $n=d=5$, we recover
the well-known sequence of sheaves as discussed in \cite{BDLR:Dq}, for
example. Given (\ref{eq:000}) and (\ref{eq:2=d}) we know the
geometrical interpretation for $(0,0,\ldots,0)_r$ for all $r$.

The case of weighted projective space is more complicated although it
is always true that
\begin{equation}
  \fbox{$(0,0,\ldots,0)_{1} \cong \O_X(1)[1]$.}
  \label{eq:RS001}
\end{equation}
Let us consider, as an example, the familiar case of weighted $\P^4$
with weights $\{2,2,2,1,1\}$ and $d=8$. It is easy to compute the
complexes of sheaves corresponding to $(0,0,\ldots,0)_{r}$ and we list
the result in table \ref{tab:wp8}.

\begin{table}
\renewcommand{\arraystretch}{1.5}
\setlength{\doublerulesep}{\arrayrulewidth}
\[\begin{array}{|c|r|}
\hline
r&\multicolumn{1}{c|}{(0,0,0,0,0)_r}\\ \hline\hline
0&\xymatrix@1{\O_X\ar[r]&0\ar[r]&0}\\ \hline
1&\xymatrix@1{\O_X(1)}\\ \hline
2&\xymatrix@1{\O_X(1)^{\oplus2}
  \ar[rr]^-{\left(\begin{smallmatrix}x_3&x_4\end{smallmatrix}\right)}&&
  \O_X(2)}\\ \hline
3&\xymatrix@1{\O_X(1)\ar[r]&\O_X(2)^{\oplus 2}\oplus\O_X(1)^{\oplus
    3}\ar[r]&\O_X(3)}\\ \hline
4&\xymatrix@1{\O_X(2)\oplus\O_X(1)^{\oplus 6}
  \ar[r]&\O_X(3)^{\oplus 2}\oplus\O_X(2)^{\oplus
    3}\ar[r]&\O_X(4)}\\ \hline
5&\xymatrix@1{\O_X(1)^{\oplus 3}\ar[r]&
  \O_X(3)\oplus\O_X(2)^{\oplus 6}\oplus\O_X(1)^{\oplus 3}
  \ar[r]&\O_X(4)^{\oplus 2}\oplus\O_X(3)^{\oplus
    3}\ar[r]&\O_X(5)}\\ \hline
6&\xymatrix@1@C=4mm{\scriptstyle\O_X(2)^{\oplus 3}\oplus\O_X(1)^{\oplus 6}\ar[r]&
  \scriptstyle\O_X(4)\oplus\O_X(3)^{\oplus 6}\oplus\O_X(2)^{\oplus 3}
  \ar[r]&\scriptstyle\O_X(5)^{\oplus 2}\oplus\O_X(4)^{\oplus
    3}\ar[r]&\scriptstyle\O_X(6)}\\ \hline
7&\xymatrix@1@C=4mm{\scriptstyle\O_X(1)^{\oplus 3}\ar[r]&
       \scriptstyle\O_X(3)^{\oplus 3}
       \oplus\O_X(2)^{\oplus 6}\oplus\O_X(1)\ar[r]&
  \scriptstyle\O_X(5)\oplus\O_X(4)^{\oplus 6}\oplus\O_X(3)^{\oplus 3}
  \ar[r]&\scriptstyle\O_X(6)^{\oplus 2}\oplus\O_X(5)^{\oplus
    3}\ar[r]&\scriptstyle\O_X(7)}\\ \hline
\end{array}\]
\caption{The case $\P^4_{\{2,2,2,1,1\}}$. The
  final term in the complex is always in position $-1$.} \label{tab:wp8}
\end{table}

Note that the complex corresponding to $(0,0\ldots,0)_2$ has
nontrivial cohomology in both positions. Thus this complex cannot
possibly be equivalent to a single sheaf. This is in contrast with the
result (\ref{eq:RS00}) for the non-weighted case.  At position $-1$,
the cohomology is equal to $\O_S(2)[1]$, where $\O_S$ is the structure
sheaf of the $\Z_2$-orbifold singularity $x_3=x_4=0$.

The sequence of complexes in table \ref{tab:wp8} shows how $\O_X$
behaves under monodromy around the \LG\ point in the moduli space as
discussed in \cite{me:navi}, for example. The
same result, using Fourier--Mukai transforms, has also been observed
recently in \cite{CK:mon}.

\subsection{General Torsion Modules} \label{ss:tors}

It is very straight-forward to build the general RS-brane
$(L_0,L_1,\ldots)_r$ from the branes $\mathbf{k}(r)\cong(0,0,\ldots)_r$ of the
preceding section. Consider the following short exact sequence:
\def\bfrac#1#2{{\displaystyle\frac{#1}{#2}}}
\begin{equation}
\xymatrix@1@M=2mm{
0\ar[r]&\bfrac{k[x]}{(x)}(-1)\ar[r]^-x&\bfrac{k[x]}{(x^2)}\ar[r]&
    \bfrac{k[x]}{(x)}\ar[r]&0.
}
\end{equation}
This shows that $k[x]/(x^2)$ is an extension coming from
$\Ext^1\Bigl(k[x]/(x),\bigl(k[x]/(x)\bigr)(-1)\Bigr)$. Using the
concept of this extension, it is straight-forward to build any RS
D-brane as a finite sequence of extensions of a collection of branes
of the type $(0,\ldots,0)_r$ considered in the previous section.

For example, we immediately have a distinguished triangle
\begin{equation}
\xymatrix@C=-10mm{
  (L_0,L_1,\ldots,L_{n-1})_0\ar[rr]|{[1]}&&(0,L_1,\ldots,L_{n-1})_{-L_0}
  \ar[ld]^(0.4)*-<2mm>{\scriptstyle x_0^{L_0+1}}\\
  &(L_0+1,L_1,\ldots,L_{n-1})_0\ar[lu]} \label{eq:triS}
\end{equation}
This observation leads to a systematic picture of the relations
between various RS D-branes. 

It is quite easy in this picture to confirm the finiteness of the
number of RS D-branes. Suppose we have a D-brane of the form
(\ref{eq:RS}) where one of the $(L_i+1)$'s is equal to $l_i$, the
degree of the corresponding variable in the superpotential $W$. Then
we may use the superpotential relation to remove this monomial from
the set of generators of the ideal in the denominator of
(\ref{eq:RS}). Thus, if we perform a free resolution of this module,
the variable $x_i$ never appears in the maps. Thus, this free
resolution can equally be performed over the ring $B$ and is
therefore finite. It follows that the module is zero in $\DSing(A)$.
This also leads to the brane-anti-brane relation
\begin{equation}
  (L_0,L_1,\ldots,L_i,\ldots,L_{n-1})_0 \cong
    (L_0,L_1,\ldots,L_{l_i-L_i-2},\ldots,L_{n-1})_{-L_i-1}[1],
\end{equation}
and so we need only consider the range $0\leq L_i\leq(l_i-2)/2$ (as
is also clear from the original construction \cite{RS:DGep}).

Another relation we obtain from triangles of the form (\ref{eq:triS}) is
\begin{equation}
  (1,0,\ldots,0)_1 = \Cone\bigl((0,0,\ldots,0)_1[-1]\to(0,0,\ldots,0)_0\bigr).
\end{equation}
Using the results of section \ref{ss:00r}, this yields an equivalence
\begin{equation}
\fbox{$(1,0,\ldots,0)_1\cong 
  \Cone\bigl(O_X(1)\to O_X[n-2]\bigr).$} \label{eq:exot}
\end{equation}
For the quintic 3-fold, with $n=5$, this result was observed in
\cite{Doug:DC} and the stability of this ``exotic'' state was
discussed in \cite{AD:Dstab}. Now we see that the relationship
(\ref{eq:exot}) is true in general for the whole class of models we
are studying in this paper.

Actually it is natural to consider the set of D-branes that can be
finitely generated from the set $\mathbf{k}(r)\cong(0,0,\ldots)_r$.
This is precisely the set of modules which are {\em torsion\/}
modules, i.e., modules which are finite-dimensional as vector
spaces. This set includes all the RS-branes but includes many more.
Let $\mathcal{S}$ denote the full subcategory of $\DC(\hbox{gr-}A)$
corresponding to bounded complexes of torsion modules. For example, the
module
\begin{equation}
  M=\frac{A}{(x_0^2,x_1^2,x_0x_1,x_2,x_3,x_4)}
\end{equation}
is in $\mathcal{S}$ but is not an RS D-brane. However, it can be built
from RS D-branes:
\begin{equation}
  M = \Cone\bigl((0,1,0,0,0)[-1]\to(0,0,0,0,0)_{-1}\big).
\end{equation}

The set $\mathcal{S}$ is of interest for the following reason. For
simplicity let us focus on the case of non-weighted projective spaces.
In that case, it is known from the work of Beilinson \cite{Bei:res}
that the set of sheaves in (\ref{eq:RS00}) form a complete exceptional
collection on the ambient $\P^{n-1}$. That is, they generate the whole
derived category of coherent sheaves on $\P^{n-1}$. Since
$\mathcal{S}$ is also precisely the category generated by these
objects, we observe that {\em the image of $\mathcal S$ in
  $\DC(\textrm{\em qgr-}A)$ is precisely the subset of D-branes on $X$ that
  arise as complexes of restrictions of sheaves on the ambient
  $\P^{n-1}$.}\footnote{The sheaves in the complex are restrictions of
  sheaves on $\P^{n-1}$ but the morphisms in the complex need not lift
  to morphisms on $\P^{n-1}$. Thus it would not be correct to assert that
  $\mathcal S$ is the image of $\DC(\P^{n-1})$ in $\DC(X)$.}


\section{Quotient Singularities} \label{s:quo}

In general, a weighted projective space has quotient
singularities. The hypersurface $W=0$ may also therefore have quotient
singularities. Thus far we have been a little careless about our
language of sheaves on $X$ when we have such singularities.

More properly we should define the quotient {\em stack\/} 
$\P$roj $A$ as $[(\Spec A\backslash{0})/k^*]$ on which it is easier to
define coherent sheaves. 

The problem is seen by a simple example copied from \cite{Cox:}. Let
us consider sheaves on the weighted projective space
$\P^2_{\{1,1,2\}}$ and let $B$ be its homogeneous coordinate ring
$k[x_0,x_1,x_2]$, where $x_2$ has degree 2. This space has a $\Z_2$-quotient
singularity at $[0,0,1]$. Consider first the module
\begin{equation}
  M = \frac{B}{(x_0,x_1)}.
\end{equation}

The stalk of the associated sheaf $\widetilde M$ at the point
$[x_0,x_1,x_2]=[a_0,a_1,a_2]$ is obtained by localizing this module on
the homogeneous ideal of functions vanishing at this point and
restricting to elements of degree zero. This localization is zero at
every point except $[0,0,1]$. The stalk at $[0,0,1]$ is $k$ and thus
this is the skyscraper sheaf of $[0,0,1]$.

Now consider
\begin{equation}
  M(1) = \frac{B}{(x_0,x_1)}(1).
\end{equation}
This is the same as $M$ except that when we localize at
$[0,0,1]$ we get zero due to the choice of gradings of the
variables. That is, $\widetilde{ M(1)}$ is the zero sheaf. So Serre's
correspondence between graded modules modulo torsion and coherent
sheaves is not strictly correct for weighted projective spaces.

Instead of coherent sheaves on $X=\Proj A$, one should consider
coherent sheaves on $\Spec A$ which are equivariant under the $k^*$
action producing the weighted projective space. In \cite{Orlov:mfc} it
was shown that the category of such objects is equivalent to
$\hbox{qgr-}A$ as required for our program.

Suppose a weighted projective space has an orbifold singularity
locally of the form $k^{n-2}/G$ for some discrete group $G\subset
\Sl(n-2,\Z)$. Then a $k^*$-equivariant sheaf on $\Spec A$ is locally
modeled by a $G$-equivariant sheaf on $k^{n-2}$. 

In the case of $\P^2_{\{1,1,2\}}$ analyzed above, it is clear how to
interpret our sheaves now. Consider the skyscraper sheaf of the origin
in $k^2$. A $\Z_2$ action on $k^2$ which fixes the origin may act as
$\pm 1$ on such a sheaf. The resulting quotient sheaf will be
associated to the modules $M$ or $M(1)$ respectively. For a longer
discussion of this and how it fits into the language of stacks we
refer, for example, to \cite{Pantev:2005wj}.

This is exactly the setting for the {\em McKay correspondence}.
Locally the McKay correspondence asserts that the derived category of
$G$-equivariant sheaves on $k^{n-2}/G$ is equivalent to the derived
category of coherent sheaves on a crepent resolution of this
singularity, assuming such a resolution exists. A necessary condition
for the existence of such a resolution is that $G\subset\Sl(n-2,\Z)$.
Mathematically the McKay correspondence has only been proven in
certain cases (see \cite{BKM:MisM}, and \cite{Bridge:ICM} for a
review). String theory essentially ``proves'' the McKay correspondence
in all cases since blowing up the orbifold is a change in $B+iJ$ which
cannot effect the B-model. Thus, B-type D-branes are unaffected by the
process. See \cite{me:TASI-D} for further details.

Let $\widetilde X$ be a smooth crepent resolution of $X$, assuming one
exists. A global form of the McKay correspondence, which we will
assume here, states an equivalence between the derived category of
sheaves on $X$ (viewed as a stack) and the derived category of sheaves
on $\widetilde X$. So, we arrive at the statement that {\em the
  category $\textrm{DGrB}(W)$ of matrix factorizations is equivalent
  to the category $\DC(\widetilde X)$.}

Actually the category $\DC(\widetilde X)$ may be constructed directly
following the construction of Cox \cite{Cox:}. $\widetilde X$ is now a
hypersurface in some toric variety $T$ obtained by blowing up the
weighted projective space. $A$ is now a multigraded algebra derived
from the homogeneous coordinates of $T$. We construct some
ideal\footnote{Denoted $B$ by Cox.} $J_\Sigma$ derived from the toric
fan $\Sigma$ associated to $T$. Then
\begin{equation}
  \DC(\widetilde X) \cong \frac{\DC(\hbox{gr-}A)}{\mathcal{J}_\Sigma},
\end{equation}
where $\mathcal{J}_\Sigma$ is the triangulated category generated by
$J_\Sigma$ and all its multigraded twists. This construction seems to
fit very naturally into Orlov's picture of equivalences between
categories but we will not pursue it further here.

\subsection{$\P^4_{\{2,2,2,1,1\}}$}  \label{ss:eg2}

Let us consider how all this works in the example
$\P^4_{\{2,2,2,1,1\}}$ studied in \cite{CDFKM:I}.  Some aspects of
D-branes and matrix factorizations for this case have already been
analyzed in \cite{Brunner:2005fv}. This model has a $\Z_2$
singularity along the curve $C$ given by $z_3=z_4=0$. $C$ is a curve
of genus 3. Let $D$ be the exceptional divisor $C\times\P^1$ one
obtains when this curve is blown up.

Let us remind ourselves of the McKay correspondence for the simplest
case $k^2/\Z_2$. Start with the skyscraper sheaf of the origin in
$k^2$. We may divide this by $\Z_2$ equivariantly in two ways
according to whether the $\Z_2$ acts as $+1$ or $-1$ on the fibre. We
denote these two $\Z_2$-equivariant sheaves $\O_0^+$ and $\O_0^-$. The
resolution of $k^2/\Z_2$ has $E=\P^1$ as the exceptional divisor. Let
$\O_E$ denote the structure sheaf of $E$ extended by zero over the
total space of the resolution. The McKay correspondence then maps
\begin{equation}
\begin{split}
  \O_0^+&\mapsto \O_E\\
  \O_0^-&\mapsto \O_E(-1)[1].
\end{split} \label{eq:McKayZ2}
\end{equation}

Extending this to the 3-fold in question, the sheaf $\O_D$,
corresponding to a 4-brane wrapping $D$, will correspond to the
$\Z_2$-invariant structure sheaf of $C$ in $\P^4_{\{2,2,2,1,1\}}$.
That is, we consider the module $M = A/(x_3,x_4)$.

We may now follow the procedure of section \ref{ss:map} to yield the
corresponding matrix factorization. We have already done step $1'$, and
step $2'$ is trivial since $M$ is already in
$\DC(\hbox{gr-}A_{\geq0})$. The nontrivial part of the analysis comes
in step $3'$. A computation with Macaulay yields
\begin{equation}
\begin{split}
  \Ext^2(\mathbf{k}(-2),M) &= k\\
  \Ext^3(\mathbf{k}(-1),M) &= k^2\\
  \Ext^4(\mathbf{k},M) &= k,
\end{split}
\end{equation}
showing that $M$ does not lie in $\mathcal{D}_0$. Instead
one builds the following complex:
\begin{equation}
\fboxsep=1pt
M''=\left(
  \xymatrix@1{\mathbf{k}[-3]^{\oplus3}\ar[rr]^-{\left(\begin{smallmatrix}x_0&x_1&x_2
    \end{smallmatrix}\right)}&&
  \mathbf{k}(-2)[-2]\ar[r]^-f&M
  \save[]+<0mm,-5mm>*{\fbox{$\scriptstyle 0$}}\restore
}\right),
\end{equation}
where $f$ generates $\Ext^2(\mathbf{k}(-2),M)$. $M''$ satisfies
$\Ext^m(\mathbf{k}(-r),M'')$ for all $m$ and all $r\geq0$ and so lies in
$\mathcal{D}_0$.  It is easy to show that $M$ has a finite free
resolution and so does not contribute to the matrix
factorization. This results in the correspondence:
\begin{equation}
\fbox{$\O_D\cong 
  \Cone\Bigl(\xymatrix@1{
    \mathbf{k}[-4]^{\oplus3}\ar[rr]^-{\left(\begin{smallmatrix}x_0&x_1&x_2
    \end{smallmatrix}\right)}&&
  \mathbf{k}(-2)[-3]}\Bigr).$} \label{eq:OD}
\end{equation}

In section \ref{s:RS} we identified the D-branes
$\mathbf{k}(r)=(0,0,\ldots,0)_r$ as specific matrix factorizations. In
our case, with $n=5$, they are given by $16\times16$ matrices. These
may be substituted into the above equation to yield an explicit
$64\times 64$ matrix factorization associated to $\O_D$. 

Now consider $M(1)$. The $\Z_2\subset k^*$ that acts as $-1$ on
$x_3$ and $x_4$, will act as $-1$ on this module. So this is the
$\Z_2$-anti-invariant sheaf supported on $C$.
It follows from (\ref{eq:McKayZ2}) and (\ref{eq:OD}) that

\begin{equation}
\fbox{$\O_D(0,-1)\cong 
  \Cone\Bigl(\xymatrix@1{
    \mathbf{k}(1)[-5]^{\oplus3}\ar[rr]^-{\left(\begin{smallmatrix}x_0&x_1&x_2
    \end{smallmatrix}\right)}&&
  \mathbf{k}(-1)[-4]}\Bigr).$} \label{eq:ODm}
\end{equation}
yielding a $64\times 64$ matrix factorization for this sheaf.

Note that we have a bigrading for the sheaf on the left of
(\ref{eq:ODm}). This is because $h^{1,1}=2$ for the resolved manifold
and so we have two directions in the K\"ahler moduli space. If we
denote a twisted sheaf by $\cF(a,b)$, then monodromy around the large
radius limit of the weighted projective space itself will increase $a$
while monodromy around the large radius limit of the exceptional set
will increase $b$. See \cite{me:navi} for more on these monodromies.

As a further example, suppose we want to find the structure sheaf of
the subspace $C\times \{p\}$ of the exceptional set $C\times\P^1$. Let
$p$ have homogeneous coordinates $[y_0,y_1]$. One can then show that
\begin{equation}
  \O_{C\times \{p\}} = \Cone\Bigl(\xymatrix@1{\O_D(0,-1)
    \ar[rr]^-{y_1x_3-y_0x_4}&&\O_D}\Bigr).
\end{equation}
Therefore this sheaf may be built by coning the above matrices to form a
factorization using $128\times 128$ matrices.


\section{Discussion} \label{s:disc}

We have demonstrated a completely systematic way of translating
between the geometric language of vector bundles or sheaves and the
language of matrix factorizations. At first sight one might consider
it an advantage to try to compute with matrix factorizations.
After all, the concept of a matrix factorization would appear to be
more straight-forward that the derived category of coherent sheaves.
However, we have discovered that the matrices for geometrical objects
rapidly become large --- we observed a $128\times128$ matrix
factorization for a 2-brane in section \ref{ss:eg2}. It is also very
awkward to compute the space of morphisms between two matrix
factorizations.  So we believe that there is no computational
advantage in using matrix factorizations over coherent sheaves.

It is interesting to observe that, for 3-folds, the complete
intersection rational curves appear to correspond to the
lowest-dimensional matrices. In the case of the quintic threefold
these $4\times 4$ matrices corresponds to the 2875 lines and 609250
quadrics. One can easily show that $2\times 2$ matrix factorizations
are impossible for a smooth 3-fold and we have not observed any
$3\times 3$ matrix factorizations.

The construction of the category of B-type D-branes in the paper has
some very striking features which appear to make it a very natural
setting for the phase picture of \CY\ manifolds. There is a single
category --- namely the derived category of graded $A$-modules that
seems to serve as the ``master'' category over the whole moduli
space. The \LG\ phase and the large radius phase have D-branes
described by particular quotient categories of this master
category. Monodromy around the limit point in each phase corresponds
to shifting the grade by one in the master category.

The semiorthogonal decompositions give a very explicit method of
taking the various quotients of the master category and allows one to
translate between the different phases with surprising ease.

Obtaining the general phase picture in toric geometry would seem to
simply be a matter of extending Orlov's results to multiply-graded
algebras. We will pursue this elsewhere.

Another obvious extension of this work involves analyzing stability
for D-branes in \LG\ theories. Although there has been some work on
this subject \cite{Wal:LGstab}, there remains much to be
discovered. There is a general belief that stability conditions should
simplify at the limit of point of each phase. Therefore there might be
a relationship between stability and Orlov's picture of semiorthogonal
decompositions.


\section*{Acknowledgments}

I wish to thank T.~Bridgeland, D.~Orlov, A.~Roy and J.~Walcher
for useful discussions. 
The author is supported by NSF grants DMS--0301476
and DMS--0606578.


\begin{thebibliography}{10}

\bibitem{W:phase}
E.~Witten,
\newblock {\em Phases of $N=2$ Theories in Two Dimensions},
\newblock Nucl. Phys. {\bf B403} (1993) 159--222, hep-th/9301042.

\bibitem{AGM:II}
P.~S. Aspinwall, B.~R. Greene, and D.~R. Morrison,
\newblock {\em \CY\ Moduli Space, Mirror Manifolds and Spacetime Topology
  Change in String Theory},
\newblock Nucl. Phys. {\bf B416} (1994) 414--480.

\bibitem{Doug:DC}
M.~R. Douglas,
\newblock {\em D-Branes, Categories and $N$=1 Supersymmetry},
\newblock J. Math. Phys. {\bf 42} (2001) 2818--2843, hep-th/0011017.

\bibitem{Doug:DICM}
M.~R. Douglas,
\newblock {\em Dirichlet Branes, Homological Mirror Symmetry, and Stability},
\newblock in ``Proceedings of the International Congress of Mathematicians:
  Beijing 2002'', vol. III, pages 395--408, Higher Education Press, Beijing,
  2003,
\newblock math.AG/0207021.

\bibitem{AL:DC}
P.~S. Aspinwall and A.~E. Lawrence,
\newblock {\em Derived Categories and Zero-Brane Stability},
\newblock JHEP {\bf 08} (2001) 004, hep-th/0104147.

\bibitem{AD:Dstab}
P.~S. Aspinwall and M.~R. Douglas,
\newblock {\em D-Brane Stability and Monodromy},
\newblock JHEP {\bf 05} (2002) 031, hep-th/0110071.

\bibitem{Gep:}
D.~Gepner,
\newblock {\em Exactly Solvable String Compactifications on Manifolds of
  $SU(N)$ Holonomy},
\newblock Phys. Lett. {\bf 199B} (1987) 380--388.

\bibitem{VW:}
C.~Vafa and N.~Warner,
\newblock {\em Catastrophes and the Classification of Conformal Theories},
\newblock Phys. Lett {\bf 218B} (1989) 51--58.

\bibitem{RS:DGep}
A.~Recknagel and V.~Schomerus,
\newblock {\em D-branes in Gepner Models},
\newblock Nucl. Phys. {\bf B531} (1998) 185--225, hep-th/9712186.

\bibitem{BDLR:Dq}
I.~Brunner, M.~R. Douglas, A.~Lawrence, and C.~R{\"o}melsberger,
\newblock {\em D-branes on the Quintic},
\newblock JHEP {\bf 08} (2000) 015, hep-th/9906200.

\bibitem{KL:Mfac}
A.~Kapustin and Y.~Li,
\newblock {\em D-branes in Landau-Ginzburg Models and Algebraic Geometry},
\newblock JHEP {\bf 12} (2003) 005, hep-th/0210296.

\bibitem{Kapustin:2003rc}
A.~Kapustin and Y.~Li,
\newblock {\em D-branes in Topological Minimal models: The Landau--Ginzburg
  Approach},
\newblock JHEP {\bf 07} (2004) 045, hep-th/0306001.

\bibitem{Brunner:2003dc}
I.~Brunner, M.~Herbst, W.~Lerche, and B.~Scheuner,
\newblock {\em Landau-Ginzburg realization of open string TFT},
\newblock hep-th/0305133.

\bibitem{Lazaroiu:2003zi}
C.~I. Lazaroiu,
\newblock {\em On the Boundary Coupling of Topological Landau--Ginzburg
  Models},
\newblock JHEP {\bf 05} (2005) 037, hep-th/0312286.

\bibitem{AADia:GepD}
S.~K. Ashok, E.~Dell'Aquila, and D.-E. Diaconescu,
\newblock {\em Fractional Branes in Landau--Ginzburg Orbifolds},
\newblock Adv. Theor. Math. Phys. {\bf 8} (2004) 461--513, hep-th/0401135.

\bibitem{Brunner:2004mt}
I.~Brunner, M.~Herbst, W.~Lerche, and J.~Walcher,
\newblock {\em Matrix Factorizations and Mirror Symmetry: The Cubic Curve},
\newblock hep-th/0408243.

\bibitem{DGJT:D}
M.~R. Douglas, S.~Govindarajan, T.~Jayaraman, and A.~Tomasiello,
\newblock {\em D-branes on Calabi--Yau Manifolds and Superpotentials},
\newblock Commun. Math. Phys. {\bf 248} (2004) 85--118, hep-th/0203173.

\bibitem{Hori:2004ja}
K.~Hori and J.~Walcher,
\newblock {\em F-term Equations Near Gepner Points},
\newblock JHEP {\bf 01} (2005) 008, hep-th/0404196.

\bibitem{HW:mfac}
K.~Hori and J.~Walcher,
\newblock {\em D-branes from Matrix Factorizations},
\newblock Comptes Rendus Physique {\bf 5} (2004) 1061--1070, hep-th/0409204.

\bibitem{Brunner:2005fv}
I.~Brunner and M.~R. Gaberdiel,
\newblock {\em Matrix Factorisations and Permutation Branes},
\newblock JHEP {\bf 07} (2005) 012, hep-th/0503207.

\bibitem{Enger:2005jk}
H.~Enger, A.~Recknagel, and D.~Roggenkamp,
\newblock {\em Permutation Branes and Linear Matrix Factorisations},
\newblock JHEP {\bf 01} (2006) 087, hep-th/0508053.

\bibitem{Orlov:mfc}
D.~Orlov,
\newblock {\em Derived Categories of Coherent Sheaves and Triangulated
  Catgeories of Singularities},
\newblock math.AG/0503632.

\bibitem{Orlov:LG}
D.~Orlov,
\newblock {\em Triangulated Categories of Singularities and D-Branes in
  Landau--Ginzburg Orbifolds},
\newblock Proc. Steklov Inst. Math. {\bf 246} (2004) 227--248, math.AG/0302304.

\bibitem{KL:TLG1}
A.~Kapustin and Y.~Li,
\newblock {\em Topological Correlators in Landau--Ginzburg Models with
  Boundaries},
\newblock Adv. Theor. Math. Phys. {\bf 7} (2004) 727--749, hep-th/0305136.

\bibitem{Wal:LGstab}
J.~Walcher,
\newblock {\em Stability of Landau--Ginzburg Branes},
\newblock J. Math. Phys. {\bf 46} (2005) 082305, hep-th/0412274.

\bibitem{Eis:mf}
D.~Eisenbud,
\newblock {\em Homological Algebra on a Complete Intersection, with an
  Application to Group Representations},
\newblock Trans. Amer. Math. Soc. {\bf 260} (1980) 35--64.

\bibitem{Serre:mp}
J.~P. Serre,
\newblock {\em Modules projectifs et espace fibr{\'e}s {\`a} fibre
  vectorielle},
\newblock S{\'e}minaire Dubreil--Pisot {\bf 23} (1958) 531--543.

\bibitem{Hartshorne:}
R.~Hartshorne,
\newblock {\em Algebraic Geometry}, Graduate Texts in Mathematics~{\bf 52},
\newblock Springer-Verlag, 1977.

\bibitem{AKH:m0}
P.~S. Aspinwall, R.~L. Karp, and R.~P. Horja,
\newblock {\em Massless D-branes on Calabi-Yau threefolds and monodromy},
\newblock Commun. Math. Phys. {\bf 259} (2005) 45--69, hep-th/0209161.

\bibitem{AM:tor}
M.~Artin and D.~Mumford,
\newblock {\em Some Elementary Results of Unirational Varieties which are not
  Rational},
\newblock Proc. London Math. Soc. {\bf 25} (1972) 75--95.

\bibitem{Eis:CA}
D.~Eisenbud,
\newblock {\em Commutative Algebra With a View Towards Algebraic Geometry},
  Graduate Texts in Mathematics~{\bf 150},
\newblock Springer, 2004.

\bibitem{me:navi}
P.~S. Aspinwall,
\newblock {\em Some Navigation Rules for D-brane Monodromy},
\newblock J. Math. Phys. {\bf 42} (2001) 5534--5552, hep-th/0102198.

\bibitem{CK:mon}
A.~Canonaco and R.~L. Karp,
\newblock {\em Derived Autoequivalences and a Weighted Beilinson Resolution},
\newblock to appear.

\bibitem{Bei:res}
A.~A. Beilinson,
\newblock {\em Coherent Sheaves on $\P^{n}$ and Problems in Linear Algebra},
\newblock Func. Anal. Appl. {\bf 12} (1978) 214--216.

\bibitem{Cox:}
D.~A. Cox,
\newblock {\em The Homogeneous Coordinate Ring of a Toric Variety},
\newblock J. Algebraic Geom. {\bf 4} (1995) 17--50, alg-geom/9210008.

\bibitem{Pantev:2005wj}
T.~Pantev and E.~Sharpe,
\newblock {\em String Compactifications on Calabi--Yau Stacks},
\newblock Nucl. Phys. {\bf B733} (2006) 233--296, hep-th/0502044.

\bibitem{BKM:MisM}
T.~Bridgeland, A.~King, and M.~Reid,
\newblock {\em Mukai implies McKay},
\newblock J. Amer. Math. Soc. {\bf 14} (2001) 535--554, math.AG/9908027.

\bibitem{Bridge:ICM}
T.~Bridgeland,
\newblock {\em Derived Categories of Coherent Sheaves},
\newblock math.AG/0602129,
\newblock to appear to in proceedings of the ICM 2006.

\bibitem{me:TASI-D}
P.~S. Aspinwall,
\newblock {\em D-Branes on Calabi--Yau Manifolds},
\newblock in J.~M. Maldacena, editor, ``Progress in String Theory. TASI 2003
  Lecture Notes'', pages 1--152, World Scientific, 2005,
\newblock hep-th/0403166.

\bibitem{CDFKM:I}
P.~Candelas et~al.,
\newblock {\em Mirror Symmetry for Two Parameter Models --- I},
\newblock Nucl. Phys. {\bf B416} (1994) 481--562, hep-th/9308083.

\end{thebibliography}

\end{document}